\documentclass{emulateapj}
\usepackage{txfonts}
\usepackage[figuresright]{rotating}

\shorttitle{BATC Photometry of NGC 188}
\shortauthors{Wang et al.}

\begin{document}
\slugcomment{AJ, in press}
\title{ BATC 15 BAND PHOTOMETRY OF THE OPEN CLUSTER NGC 188}
\author{Jiaxin Wang\altaffilmark{1,2}, Jun Ma\altaffilmark{1}, Zhenyu Wu\altaffilmark{1}, Song Wang\altaffilmark{1}, Xu Zhou\altaffilmark{1}}
\altaffiltext{1}{Key Laboratory of Optical Astronomy, National Astronomical Observatories, Chinese Academy of Sciences, Beijing 100012, China; majun@nao.cas.cn}
\altaffiltext{2}{University of Chinese Academy of Sciences, Beijing 10039, China}

\begin{abstract}
This paper presents CCD multicolour photometry for the old open cluster NGC 188. The observations were carried out as a part of the Beijing--Arizona--Taiwan--Connecticut Multicolour Sky Survey from 1995 February to 2008 March, using 15 intermediate-band filters covering 3000--10000 \AA. By fitting the Padova theoretical isochrones to our data, the fundamental parameters of this cluster are derived: an age of $t=7.5\pm 0.5$ Gyr, a distant modulus of $(m-M)_0=11.17\pm0.08$, and a reddening of $E(B-V)=0.036\pm0.010$. The radial surface density profile of NGC 188 is obtained by star count. By fitting the King model, the structural parameters of NGC 188 are derived: a core radius of $R_{c}=3.80'$, a tidal radius of $R_{t}=44.78'$, and a concentration parameter of $C_{0}=\log(R_{t}/R_{c})=1.07$. Fitting the mass function to a power-law function $\phi(m) \propto m^{\alpha}$, the slopes of mass functions for different spatial regions are derived. We find that NGC 188 presents a slope break in the mass function. The break mass is $m_{\rm break}=0.885~M_{\odot}$. In the mass range above $m_{\rm break}$, the slope of the overall region is $\alpha=-0.76$. The slope of the core region is $\alpha=1.09$, and the slopes of the external regions are $\alpha=-0.86$ and $\alpha=-2.15$, respectively. In the mass range below $m_{\rm break}$, these slopes are $\alpha=0.12$, $\alpha=4.91$, $\alpha=1.33$, and $\alpha=-1.09$, respectively. The mass segregation in NGC 188 is reflected in the obvious variation of the slopes in different spatial regions of this cluster.
\end{abstract}


\section{Introduction}

Open clusters are important tools to investigate the Galaxy evolution and stellar evolution. Due to their accurate parameters (e.g. age, metallicity, and distance), open clusters can probe the Galactic disk and trace the Galaxy evolution. For example, \citet{chen03} compiled the most complete open cluster sample with metallicity, age, and distance data, as well as kinematic information, available, and presented some statistical analysis on spatial and metallicity distributions. Their result supports the inside-out Galactic disk formation mechanism, in which the invoked star formation rate and infall timescale vary with radius. \citet{bonatto06} used a sample consisting of 654 open clusters to determine the thin-disk scale height, displacement of the Sun above the Galactic plane, scale length and the open cluster age-distribution function. The dynamical state of open clusters is complexly determined by the primordial gas expulsion, stellar evolution, dynamics, and the effect of Galactic tidal field \citep{kroupa01a,wu07,bonatto05b}: (1) the primordial expulsion of gas unbinds most of cluster, but a significant fraction of it can condense by two-body interactions to became an open cluster; (2) stellar evolution at the early phase causes large mass loss that will result in the disruption of open clusters; (3) internal dynamical evolution will result in mass segregation and evaporation of low mass star, and (4) the Galactic tidal force will strip the stars from the open cluster. Therefore, the detailed study of the dynamical state of open clusters can examine the stellar evolution theory and Galactic dynamics as well as map the Galactic tidal field. The accurate and precise fundamental parameters, structural parameters and mass function are the cornerstone for these investigations.

The photometric system of the Beijing--Arizona--Taiwan--Connecticut (BATC) Multicolour Survey consists 15 intermediate-band filters which cover the wavelength range $3000-10000~\rm{\AA}$. This system is designed to avoid most of the known bright and variable night-sky emission lines, and it is suitable to derive spectral energy distributions (SEDs) of objects. Before February 2006, a Ford Aerospace $2{\rm k}\times2{\rm k}$ thick CCD camera was applied, which has a pixel size of 15 $\mu\rm{m}$ and a field of view of $58^{\prime} \times 58^{\prime}$, resulting in a resolution of $1''.67~\rm{pixel}^{-1}$. After February 2006, a new $4{\rm k}\times4{\rm k}$ CCD with a pixel size of 12 $\mu$m was used with a resolution of $1''.36$ pixel$^{-1}$ \citep{fan09}. Due to the large field of view and deep photometry of the BATC photometric system, it is suitable for investigating open clusters in the Milky Way. The BATC photometric system has been used to study the membership determination, fundamental parameters, structural parameters and mass function of open clusters such as M67, M48, and NGC 7789 \citep{fan96,wu05,wu06,wu07}. In this paper we will study an old open cluster NGC 188 based on the BATC photometric data. We will derive the fundamental parameters and structural parameters, and discuss the mass function and dynamical state for this open cluster.

The old open cluster NGC 188 has once been regarded as the oldest open cluster in the Milky Way. Old open clusters are important to understand the early history of the Milky Way. Owing to its special location $l=122.85^{\circ}$ and $b=+22.39^{\circ}$, NGC 188 is less contaminated and easy to be observed. Therefore, since from \citet{sandage62} many works about NGC 188 have been done.  Works about the fundamental parameters of this cluster are listed in Table 1 of \citet{fornal07} and Table 1 of this paper. From these tables we can see that many methods are used to derive the fundamental parameters, and the results are somewhat different. The age of NGC 188 is from 4.3 Gyr to 16 Gyr, the metallicity is from $\rm{[Fe/H]}=-0.6$ to $ \rm{[Fe/H]}=0.69$, the reddening value is from $E(B-V)=0$ to $E(B-V)=0.15$, and the distance is from 1275 pc to 2047 pc. We consider that, one hand, these scatters result from the reddening-metallicity degeneracy. On the other hand, it reflects that different stellar evolutionary models and different photometric data will derive different results. In addition, quite some works adopted solar metallicity to NGC 188 and derived the other parameters based on this metallicity. Based on the BATC photometric data, we will try to find a method to avoid the reddening-metallicity degeneracy. The color-color diagram based on the photometric data in the four BATC bands performs well to this purpose for NGC 188. \citet{bonatto05b} has derived the structural parameters and mass function (MF), and studied the dynamical evolution of NGC 188 in detail. In addition, till now only \citet{bonatto05a} and \citet{bonatto05b} studied the MF break of NGC 188, however, because of not deep photometry, \citet{bonatto05a} and \citet{bonatto05b} did not observe the MF break of NGC 188. So, in this paper, we will study the dynamical evolution of NGC 188 based on our deeper photometry.

We describe our new observational data and photometric reduction of NGC 188 in Section 2. In Section 3 we derive the fundamental parameters of NGC 188. The structure of NGC 188 is discussed in Section 4. The MF and the mass segregation are discussed in Section 5. A summary is presented in Section 6.

\section{Observation and photometric data }

\begin{figure*}[htb]
\center
{\includegraphics[width=0.5\textwidth]{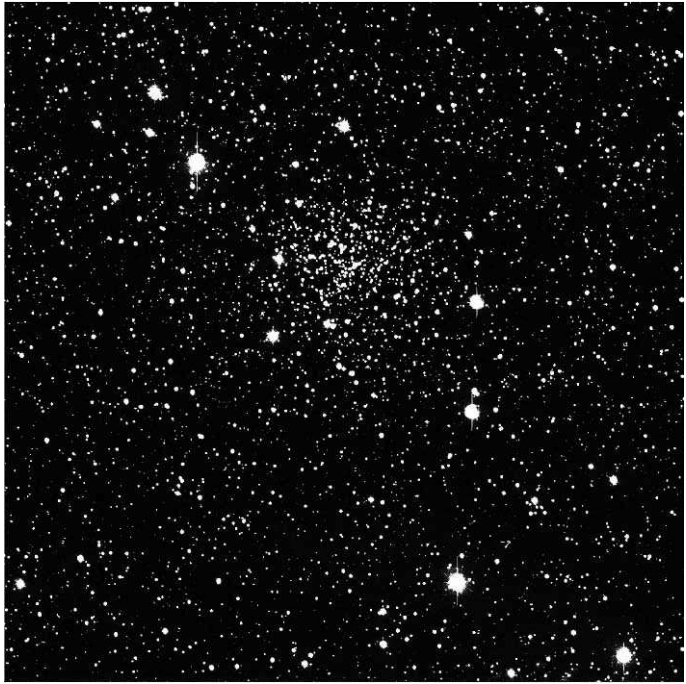}}
\caption{An image in the field of NGC 188 observed by the BATC $i$ band. The image size is $58' \times 58'$.}\label{fig1}
\end{figure*}

\begin{figure*}
\centerline{\includegraphics[width=0.5\textwidth]{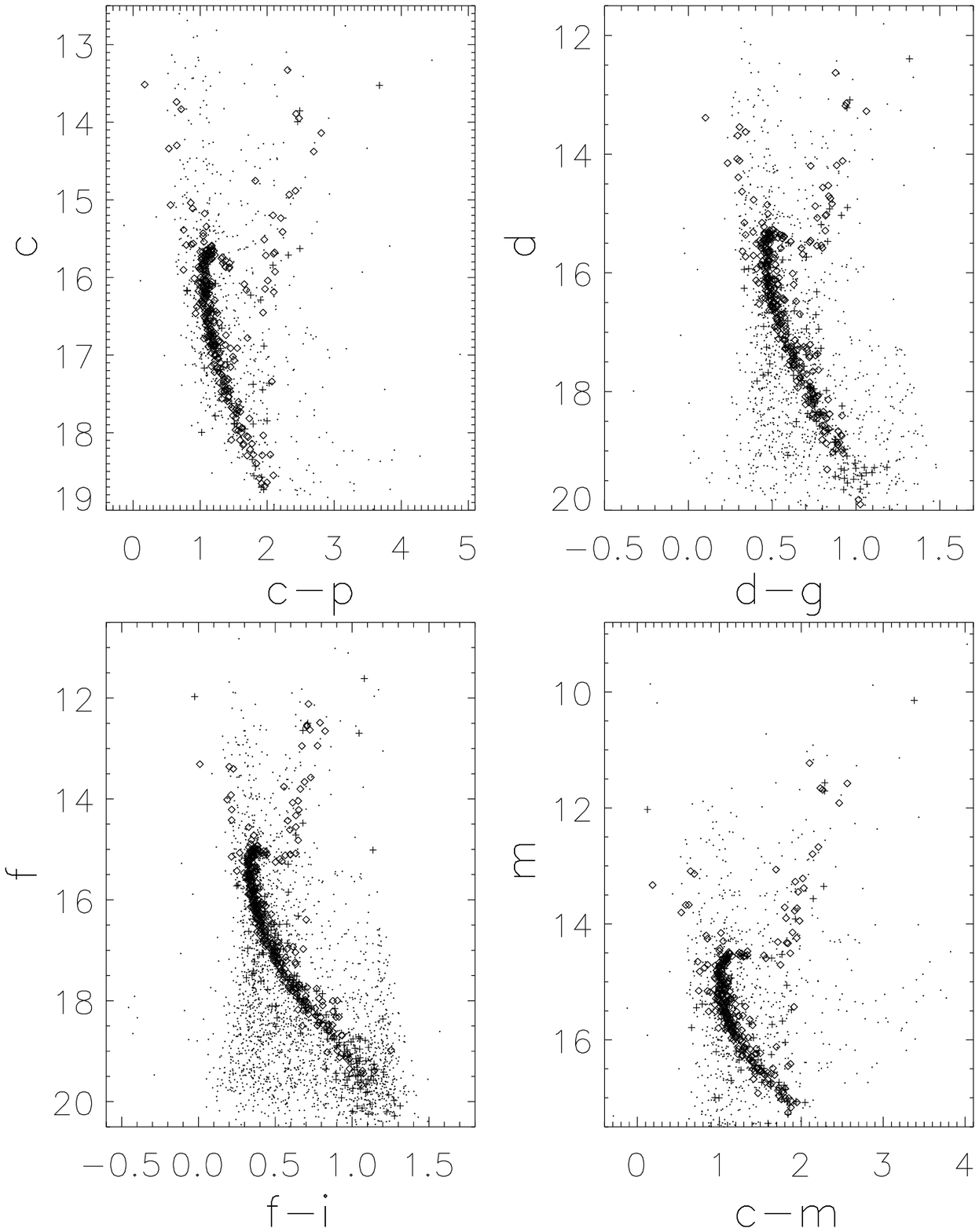}}
\caption{Observed $(c-p)$ vs. $c$, $(d-g)$ vs. $d$, $(f-i)$ vs. $f$, and $(c-m)$ vs. $m$ CMDs of NGC 188. Dots represent the stars with the membership probability of $0\% \leq P < 60\%$, crosses represent the stars with the membership probability of $60\% \leq P < 90\%$, and diamonds represent the stars with the membership probability of $90\% \leq P \leq 100\%$.}
\label{fig2}
\end{figure*}

Our observations are based on the BATC photometric system. The BATC photometric system utilizes the 60/90 cm f/3 Schmidt telescope which is at the Xinglong Station of the National Astronomical Observatories, China Academy of Sciences (NAOC). Figure 1 shows the image of NGC 188 in the BATC $i$ band. The details of the BATC system are described in \citet{yan00} and \citet{zhou01}. In Table 2, we list the corresponding effective wavelength and full width at half maximum (FWHM) of each filter, and exposure time and the number of frames observed in each filter for our observation.

Data reductions are carried out by two procedures called PIPELINE I and PIPELINE II. These procedures are developed as standard for the BATC photometric system. Bias subtraction and field flatting are carried out by PIPELINE I. Instrumental magnitudes of point sources are determined by PIPELINE II. After that, the instrumental magnitudes are calibrated to the BATC standard system \citep{zhou03}. For each filter, the average calibration error is below 0.02 mag. For every image, the mean of the FWHM of the point spread function (PSF) is $4.39''$, which critically samples ($\approx 2.5$ pixels) the PSF. So, the effect of undersampling on our photometric accuracy is not critical. For each star observed more than once in a BATC band, the final photometric result in that band is the weighted mean.

\section{Fundamental parameters for NGC 188}

We adopt the proper-motion membership probabilities in Table 3 of \citet{platais03} for NGC 188. Matching this table and our star catalog, we pick up the common stars, which are plotted in Figure 2. In this paper, stars whose proper-motions membership probabilities greater than $60\%$ are used to derive the fundamental parameters. We used the photometric data in the seven BATC bands to construct four color-magnitude diagrams (CMDs). These combinations are $(c-p)$ vs. $c$, $(d-g)$ vs. $d$, $(f-i)$ vs. $f$, and $(c-m)$ vs. $m$. We fit the theoretical BATC isochrones (see Section 3.2) to our observational data of NGC 188 to derive the fundamental parameters.

\subsection{Isochrones of stellar evolutionary models}

Stellar evolutionary models from the Padova group \citep[][and references therein]{bertelli94,Girardi00,Girardi02} are widely used. In the Padova stellar evolutionary models, \citet{Girardi02} provided tables of theoretical isochrones in such photometric systems as ABmag, STmag, VEGAmag, and a standard star system. The complete data-base \citep{Girardi02} covers a very large range of stellar masses (typically from 0.6 to $120~M_{\odot}$). These models \citep{Girardi02,Girardi08} are computed with updated opacities and equations of state, and moderate amount of convective overshoot. However, the isochrones are presented for only 6 initial chemical compositions: ${\rm [Fe/H]}=-2.2490$, $-1.6464$, $-0.6392$, $-0.3300$, $+0.0932$ (solar metallicity), and $+0.5595$, which are evidently not dense enough. It is fortunate that \citet{Marigo08} provide tables for any intermediate value of age and metallicity via an interactive web interface \footnote{http://stev.oapd.inaf.it/cmd}.

\subsection{Isochrone fitting}

To determine the main characteristics of the population in NGC 188, we fit isochrones to the cluster CMD. We used the Padova theoretical isochrones in the BATC system \citep{Marigo08}. Via an interactive web interface (see Section 3.1), we can construct a grid of isochrones for different values of age and metallicity, photometric system, and dust properties. We use the default models that involve scaled solar abundance ratios (i.e., $[\rm{\alpha/Fe]}=0.0$). In performing, the Chabrier lognormal initial mass function (IMF) \citep{chabrier01} is adopted.

\subsubsection{Reddening}

To determine the reddening value of an open cluster, a method based on the color-color diagram fitting is usually used. However, in this method, the degeneracy between reddening and metallicity requires that the metallicity is previously known. For instance, adopting the metallicity of ${\rm [Fe/H]=-0.04\pm0.05}$ for NGC 188, \citet{sara99} obtained NGC 188's reddening to be $E(B-V)=0.09 \pm 0.02$ based on the $UBV$ color-color diagram and the Hyades fiducial sequence. Assuming a solar metallicity for NGC 188, \citet{fornal07} obtained NGC 188's reddening to be $E(B-V)=0.025 \pm 0.005$ based on the $(g'-r')$ and $(u'-g')$ color-color diagram and the \citet{Girardi04} SDSS isochrones.

We derive the reddening of NGC 188. Table 1 of  \citet{fornal07} and Table 1 of this paper list the metallicities of NGC 188 obtained from spectral analysis, which are between $\rm [Fe/H]=-0.12$ and $\rm [Fe/H]=+0.12$. We adopted these two values of metallicity to derive the reddening of NGC 188 on the $(d-g)$ and $(f-i)$ color-color diagram. In addition, the resent results showed that the age of NGC 188 is between 6.0 and 8.0 Gyr (see Section 3.3 for details). In this paper, the reddening $E(B-V)$ is transformed to each BATC band using the extinction coefficient derived by \citet{chen00} based on the procedure given in Appendix B of \citet{schl98}.

Figure 3 shows the result obtained for three values of reddening $E(B-V)=0.0$, 0.036, and 0.082 ($\rm [Fe/H]=-0.12$). In the left and right panels, the BATC isochrones are for 6.0 and 8.0 Gyr, respectively. In Figure 3, the two vertical dashed lines represent the main-sequence band between $(d-g)=0.65$ and 1.05, which is the region where the data and isochrones should fit best. From Figure 3, we can see that the best fit of the BATC isochrone to the data points was obtained for $E(B-V)=0.036$. Figure 4 shows the result as Figure 3 did, but for $\rm [Fe/H]=+0.12$. From Figure 4, we can also see that the best fit of the BATC isochrone to the data points was obtained for $E(B-V)=0.036$. As a result, we arrived at a reddening value of $E(B-V)=0.036 \pm 0.010$ for NGC 188, where the quoted uncertainty only includes an estimate of the error in the reddening determination.

\begin{figure*}[htb]
\begin{center}
\centerline{\includegraphics[width=0.7\textwidth]{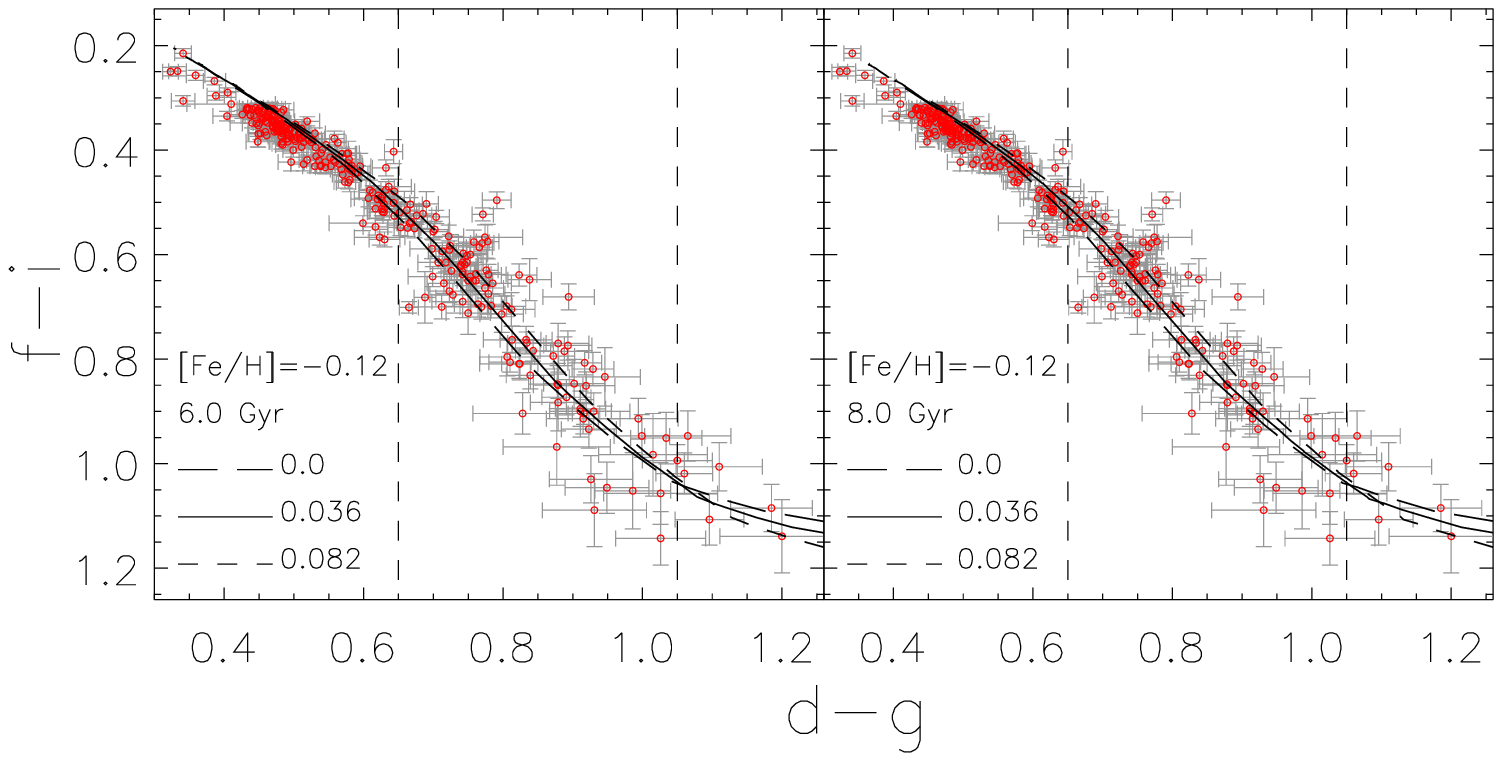}}
\caption{Observed $d-g$ vs. $f-i$ color-color diagram for the main-sequence stars with membership probability larger than $60\%$ in NGC 188. The dotted, solid, and dashed lines are BATC isochrones ($\rm [Fe/H]=-0.12$) with three different reddening values $E(B-V)=0.0$, 0.036, and 0.082. In the left and right panels, the BATC isochrones are for 6.0 and 8.0 Gyr, respectively. The two vertical dashed lines represent the main-sequence band between $(d-g)=0.65$ and 1.05, which is the region where the data and isochrones should fit best. Error bars represent the photometric errors.}\label{fig3}
\end{center}
\end{figure*}

\begin{figure*}[htb]
\center
\centerline{\includegraphics[width=0.7\textwidth]{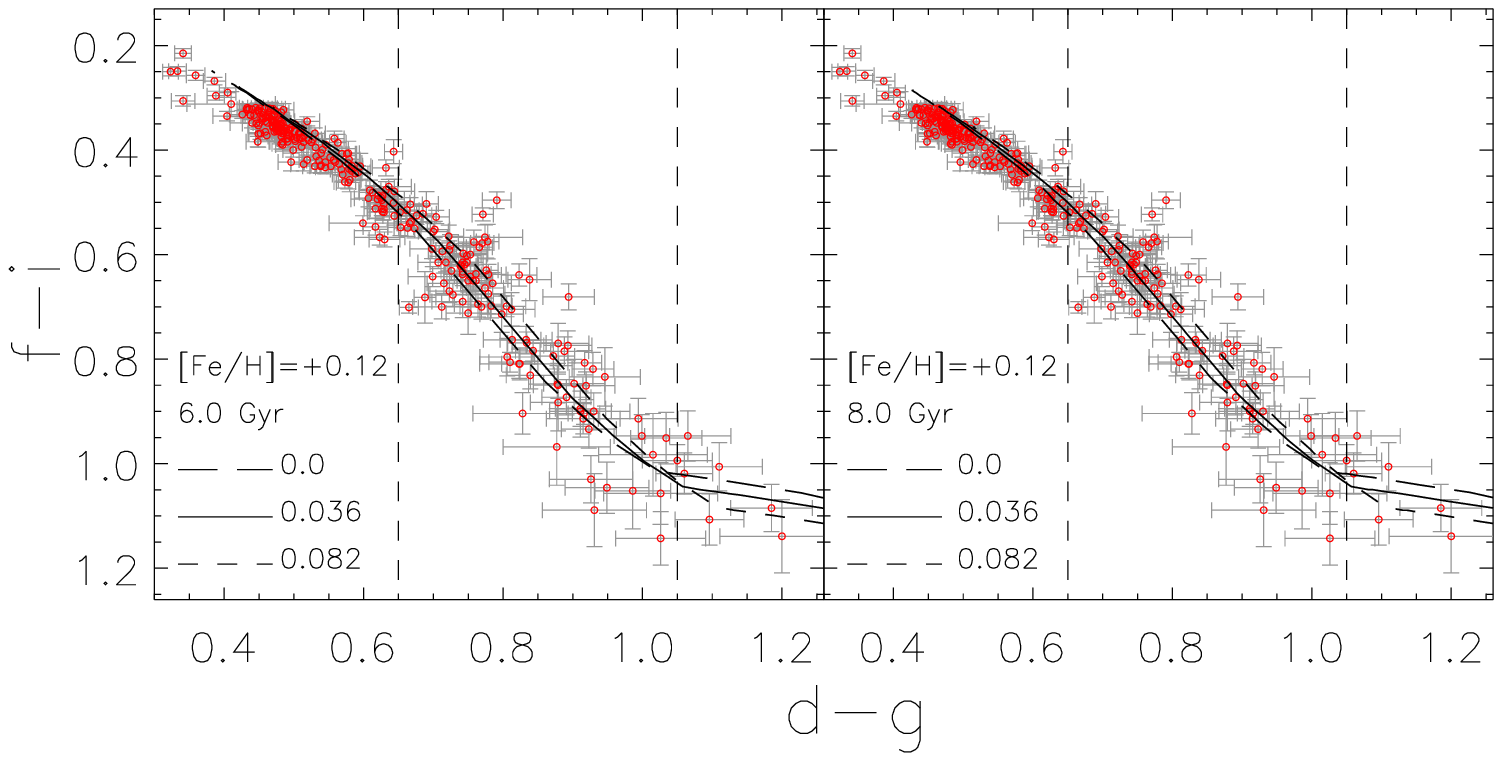}}
\caption{Observed $d-g$ vs. $f-i$ color-color diagram for the main-sequence stars with membership probability larger than $60\%$ in NGC 188. The dotted, solid, and dashed lines are BATC isochrones ($\rm [Fe/H]=+0.12$) with three different reddening values $E(B-V)=0.0$, 0.036, and 0.082. In the left and right panels, the BATC isochrones are for 6.0 and 8.0 Gyr, respectively. The two vertical dashed lines represent the main-sequence band between $(d-g)=0.65$ and 1.05, which is the region where the data and isochrones should fit best. Error bars represent the photometric errors.}\label{fig4}
\end{figure*}

\subsubsection{Metallicity, age, and distance modulus}

Since the value of reddening has been fixed, we go ahead to determine the other parameters of NGC 188 in the CMDs. Because of the age-metallicity degeneracy, we cannot simultaneously determine age and metallicity by the fitting method. We adopt a value of metallicity $~\rm [Fe/H]=0.12$ in \citet{friel10}, which was obtained from spectral analysis of red giants of NGC 188. Now, fitting the theoretical isochrone to our color-magnitude data points became straightforward. First, we chose the age for which the fitted isochrone had the same color of the turn-off as our data points. Then, by shifting the isochrone up and down (changing the distance modulus), we finally obtained the best fit.

Based on the age obtained by the previous works (see Table 1 of this paper and Table 1 of Fornal et al. 2007 in detail), we used the interactive web in Section 3.1 to construct a fine grid of isochrones about ages, sampling an age range $6.0 \leq t \leq 8.0$ Gyr at intervals of 0.1 Gyr. The metallicity ${\rm[Fe/H]} = \log(Z/Z_\odot) = 0.12$ where $Z_\odot \approx 0.0152$, so this metallicity corresponds to $Z=0.0200$. Using this fine grid of isochrones, we obtained the fundamental parameters of NGC 188. A set of the best fitting parameters for our data is adopted as follows: an age of $t=7.5\pm0.5$ Gyr, and a distance modulus of $(m-M)_{0}=11.17\pm 0.08$. The error of the age of NGC 188 is estimated by eye in the CMDs. Figure 5 shows a set of CMDs in the BATC system for NGC 188. From Figure 5, we can give an estimate of the error of age to be $\pm 0.5$ Gyr. The error of the distance modulus is also estimated by eye in the CMDs. Figure 6 shows a set of CMDs in the BATC system for NGC 188. From Figure 6, we can give an estimate of the error of distance modulus to be $\pm 0.08$.

\begin{figure*}[htb]
\center
\centerline{\includegraphics[width=0.5\textwidth]{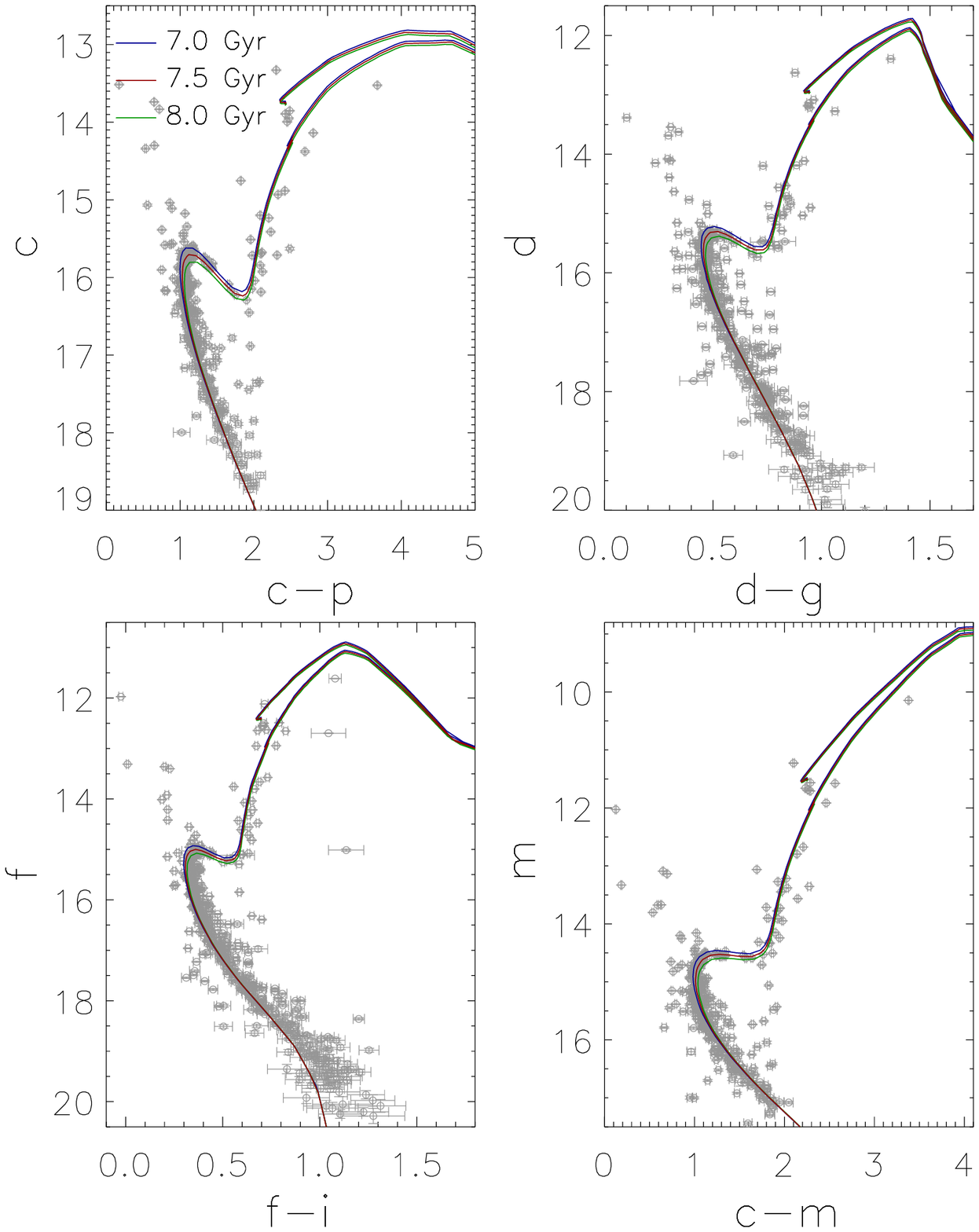}}
\caption{Observed $(c-p)$ vs. $c$, $(d-g)$ vs. $d$, $(f-i)$ vs. $f$, and $(c-m)$ vs. $m$ CMDs of NGC 188 for stars with membership probability larger than $60\%$. The blue, red, and green lines are BATC isochrones ($Z=0.0200$, $(m-M)_0=11.17$, and $E(B-V)=0.036$) with three different ages $t=7.0$ Gyr, 7.5 Gyr, and 8.0 Gyr. Error bars represent the photometric errors.}\label{fig5}
\end{figure*}

\begin{figure*}[htb]
\center
\centerline{\includegraphics[width=0.5\textwidth]{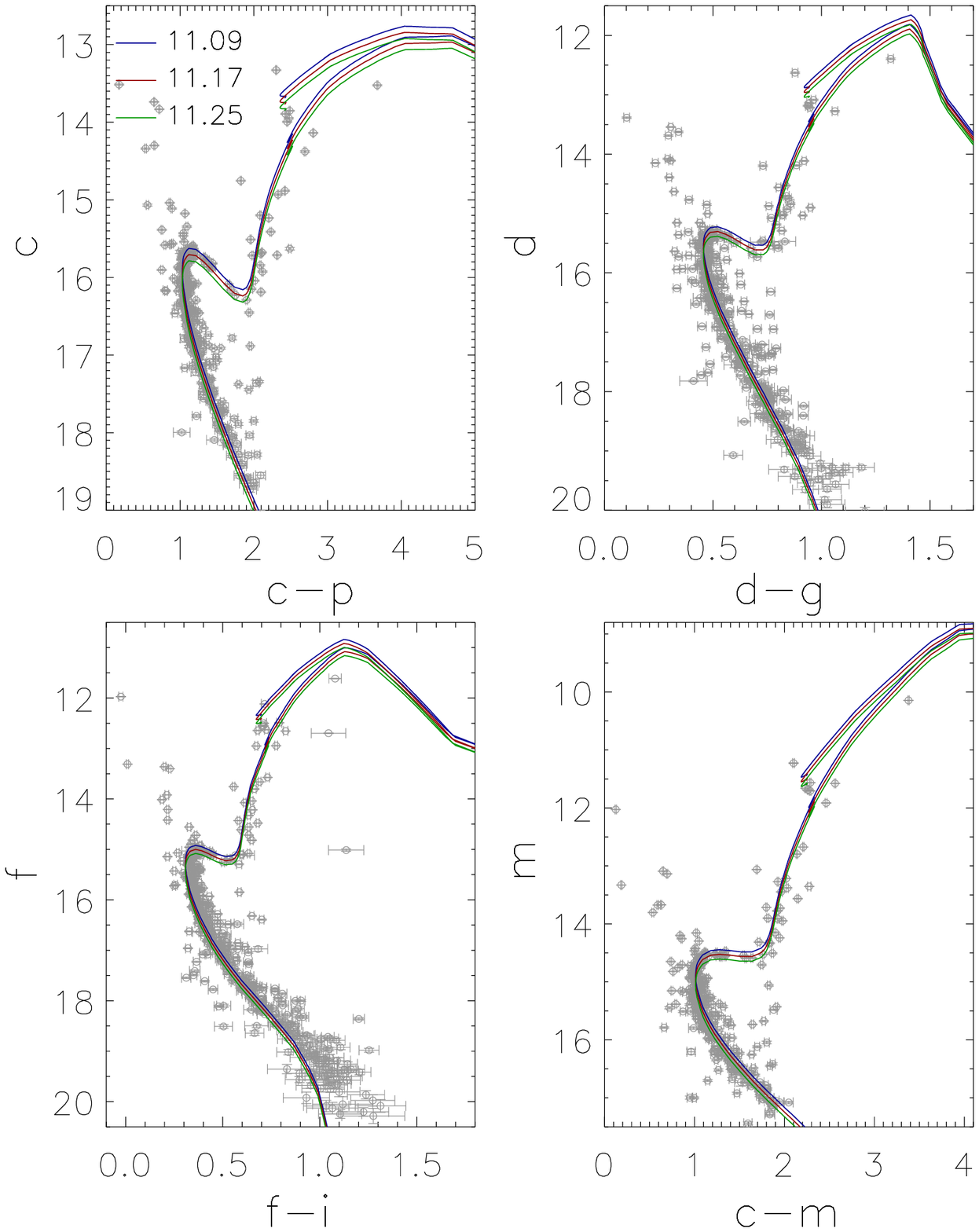}}
\caption{Observed $(c-p)$ vs. $c$, $(d-g)$ vs. $d$, $(f-i)$ vs. $f$, and $(c-m)$ vs. $m$ CMDs of NGC 188 for stars with membership probability larger than $60\%$. The blue, red, and green lines are BATC isochrones ($t=7.5$ Gyr, $Z=0.0200$, and $E(B-V)=0.036$) with three different distance moduli $(m-M)_0=11.09$, 11.17, and 11.25. Error bars represent the photometric errors.}\label{fig6}
\end{figure*}

\subsection{Comparing with previous works}

We compare our best fitting parameters for NGC 188 with those of previous studies in the past two decade. For reddening $E(B-V)$, many works (see Table 1 of Fornal et al. 2007) obtained the values of reddening to be larger than $0.08$. Our best fitting result of $0.036\pm0.010$ is in very agreement with those in \citet{carraro94}, who obtained the values of reddening to be 0.03 and 0.04. Our result is also consistent with that of \citet{fornal07}, who obtained the value of reddening to be $0.025\pm0.005$. For age, our best fitting of $7.5\pm 0.5$ Gyr is consistent with those of \citet{carraro94}, \citet{sara99}, and \citet{fornal07}, who obtained the ages of NGC 188 to be 7.5 Gyr and 7.0 Gyr, $7.0\pm0.5$ Gyr, and $7.5\pm0.7$ Gyr, respectively. For distance modulus $(m-M)_{0}$, our best fitting of $11.17\pm0.08$ (or $1714\pm64~\rm pc$) is consistent with those of \citet{carraro94}, \citet{sara99}, and \citet{fornal07}, who obtained the distances of NGC 188 to be $1705~\rm pc$ and $1720~\rm pc$, $1710\pm70~\rm pc$, and $1700\pm100~\rm pc$, respectively.

\section{Structural parameters of NGC 188}

The structure of a cluster is described by means of the radial surface density profile, which is defined as the projected number of stars per area in the direction of the cluster. In this paper we adopt the empirical formula derived by \citet{king62} to fit the radial surface density profile of NGC 188.

In order to enlarge the number of member stars of NGC 188, we used the SED fitting method developed by \citet{wu06} to search more member stars. For the $j$th star, a parameter $S$ can be defined:\\

$S_{j}(t,Z,(m-M)_{0},E(B-V))$
\begin{equation}
=\sum\limits_{i=1}^{n}\frac{[m_{ij}-M_{i}(t,Z,(m-M)_{0},E(B-V))]^2}{\sigma_{ij}^2}
\label{eq1}
\end{equation}
where $M_{i}(t,Z,(m-M)_{0},E(B-V))$ is the theoretical magnitude in the $i$th BATC band, corrected by the distance modulus $(m-M)_{0}$ and reddening $E(B-V)$ and computed from the chosen theoretical isochrone model with the age $t$ and metallicity $Z$. The values of $(m-M)_{0}$, $E(B-V)$, $t$, and $Z$ have been obtained in Section 3. Here $m_{ij}$ and $\sigma_{ij}$ are the observed magnitude and its error, respectively, of the $j$th star in the $i$th band, and $n$ is the total number of observed bands for the $j$th star. For $M_{i}$ with different stellar masses, the minimum of $S_j$, $S_{j,min}$, can be obtained for the $j$th star with the chosen theoretical model. If the observed SEDs can match the theoretical SEDs, the parameter $S_{min}$ should be the $\chi^2$ distribution with $n-P$ degrees of freedom, where $P$ is the number of free parameters to be solved. The integral probability at least as large as $S_{j,min}$ in the $\chi^2$ distribution with $n-P$ degrees of freedom is taken as the ``photometric`` membership probability of the $j$th star \citep{wu06}. We determine 2131 stars to be photometric members with $P_{\rm phot}$ greater than $50\%$. The $d-g$ versus $d$ CMD of these stars are shown in Figure 7.

\begin{figure*}[htb]
\center
\centerline{\includegraphics[width=0.4\textwidth]{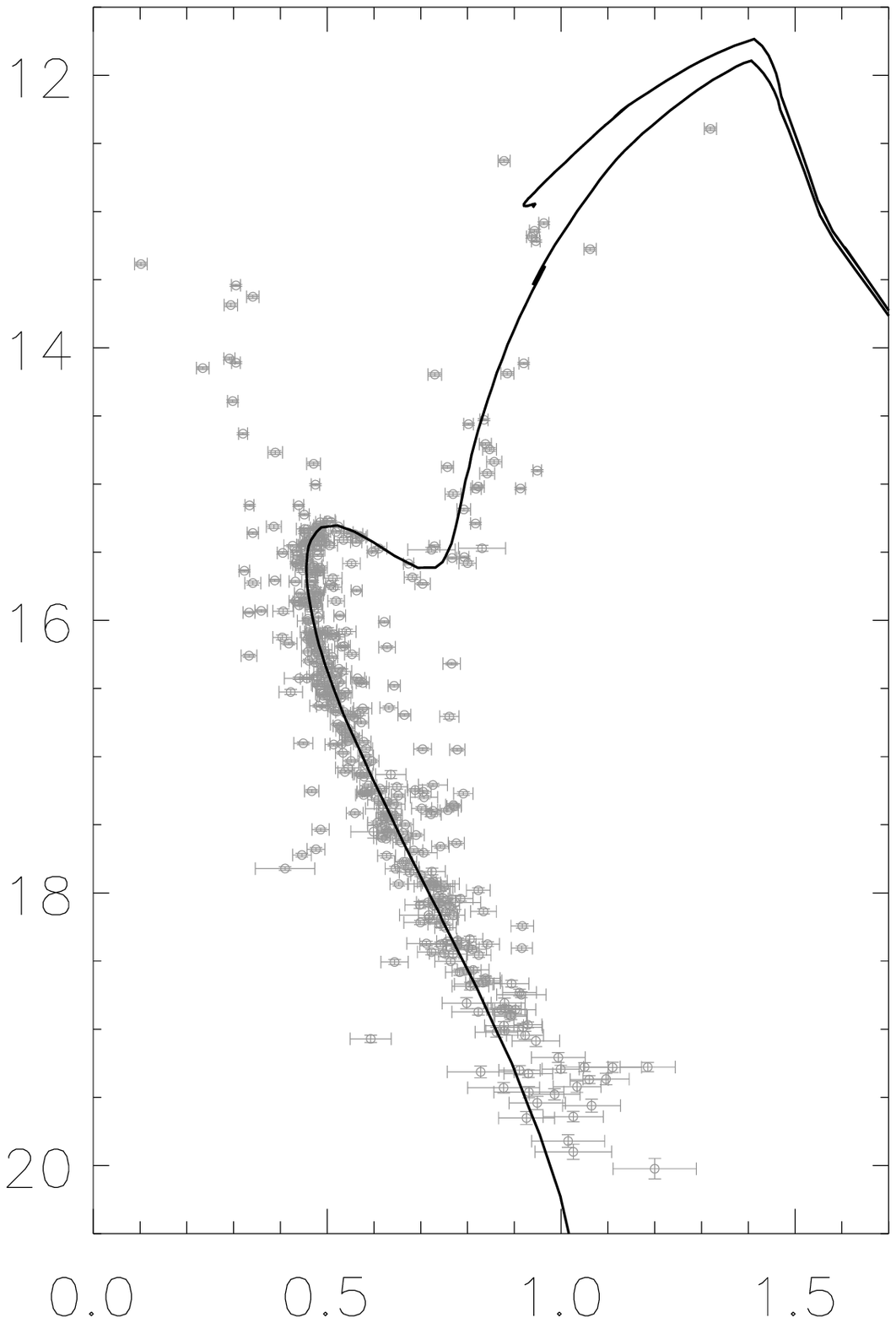}}
\caption{Observed $d-g$ vs $d$ CMD of NGC 188 for member stars which are determined by the SED fitting method. Error bars represent the photometric errors.}\label{fig7}
\end{figure*}

We used these 2131 member stars to determine the radial surface density profile of NGC 188. We adopted the central position for NGC 188 at $\alpha_0= 00^{\rm h}47^{\rm m}53^{\rm s}$ and $\delta_0 = +85^{\circ}15'30''$ (J2000.0) following \citet{bonatto05b}. The radial surface density profile is determined by counting stars inside concentric annuli with a step of $1.0'$ in radius of $0' \leq r < 20' $ and counting stars inside concentric annuli with a step of $2.0'$ in radius of $20' < r \leq 30'$. The radial surface density profile of NGC 188 is shown in Figure 8. The fit has been performed using a non-linear least-square fit routine which uses the 1 $\sigma$ Poisson errors as weights. The fitted structural parameters obtained here are as follows: a core radius of $R_{c}=3.80'$, a tidal radius of $R_{t}=44.78'$, and a concentration parameter of $C_{0}=\rm log$ $(R_{t}/R_{c})=1.07$. With the distance to the Sun $d_{\odot}=1714~\rm pc$ obtained in Section 3, the core radius and tidal radius of NGC 188 turn out to be $R_{c}=1.89~\rm pc$ and $R_{t}=22.33~\rm pc$, respectively. The fitting curves are shown in Figure 8 with solid lines. As shown in Figure 8, the radial surface density profile is reproduced well by the King model within uncertainties.

We compare our structural parameters for NGC 188 with those of previous studies. For the core radius $R_{c}$, our best fitting result of $3.80'$ is somewhat larger than that of \citet{bonatto05b}, who obtained the value of core radius to be $3.1\pm0.2'$. However, our result is somewhat smaller than that of \citet{chumak10}, who obtained the value of core radius to be $4.45'$. The tidal radius of $R_{t}=44.78'$ obtained here is somewhat larger than that in \citet{keenan73}, who obtained the tidal radius of NGC 188 to be $27.0\pm5.0'$. Our result is consistent with that in \citet{bonatto05b}, who obtained the tidal radius of NGC 188 to be $44\pm9'$. And our result is smaller than that in \citet{chumak10}, who derived the tidal radius of NGC 188 to be $71.6'$.

 \begin{figure*}[htb]
\center
\centerline{\includegraphics[width=0.5\textwidth]{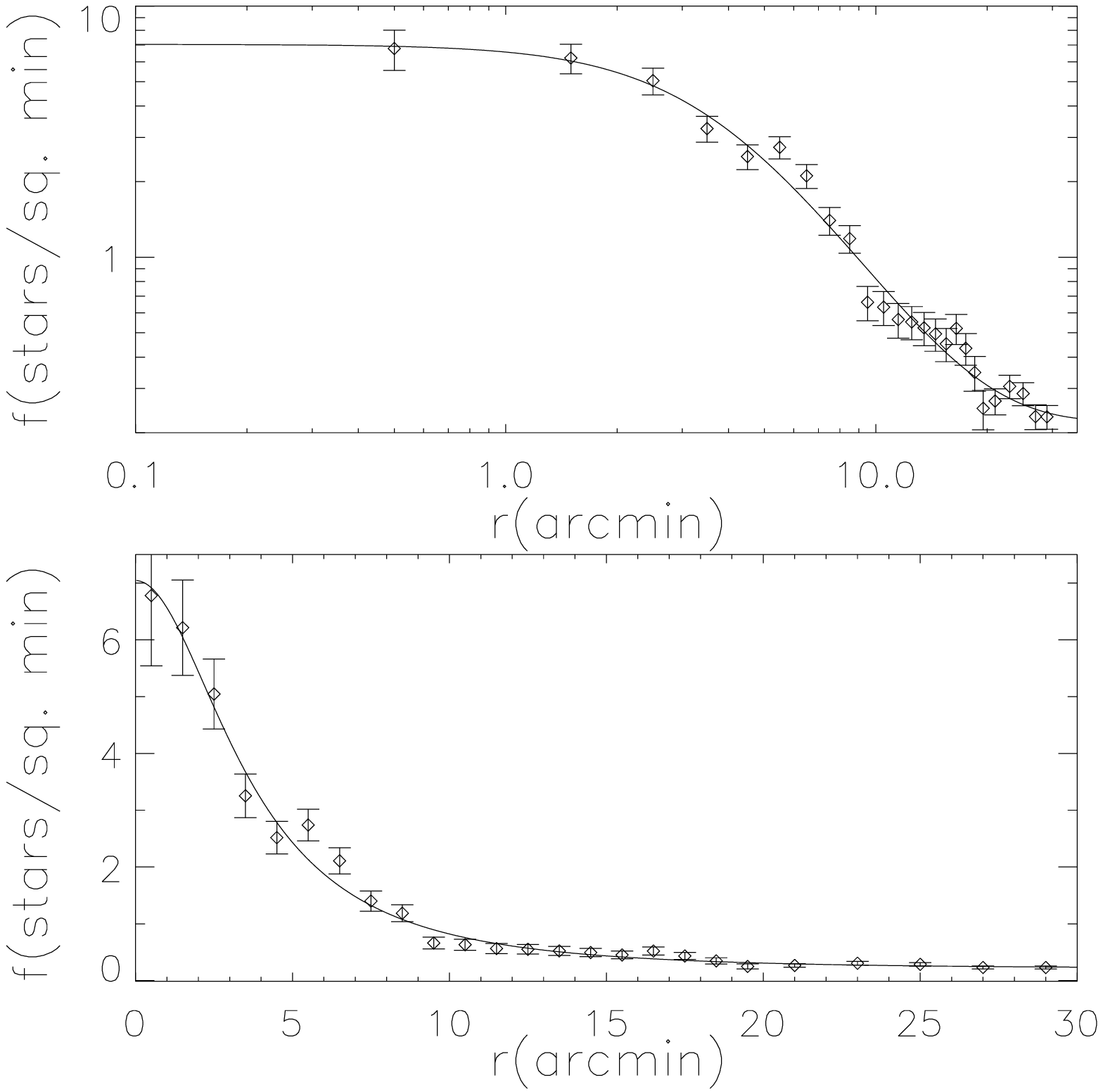}}
\caption{Radial surface density profile of NGC 188. The top panel uses logarithm coordinate, and the bottom panel uses linear coordinate. Error bars represent the 1 $\sigma$ Poisson errors. The fitting parameters of King model are adopted as follows: a core radius of $R_{c}=3.80'$ and a tidal radius of $R_{t}=44.78'$.}\label{fig8}
\end{figure*}

\section{Mass function of NGC 188}

Since a large field of view of NGC 188 is observed here, we have an opportunity to study the MF and the mass segregation for NGC 188. The member stars of NGC 188 obtained here by the SED fitting method is used. In addition, we take the stars below the main-sequence turn-off to derive MF. We used stars with $P_{\rm phot}$ greater than $60\%$ to avoid the field-star contamination. In order to study the mass segregation of NGC 188, we must know the mass functions in different regions. We select the core region ($0'\leq R <3.80'$), $3.80'\leq R <15'$, $15'\leq R <30'$, and the whole region ($0'\leq R \leq30'$). Completeness estimation is necessary to obtain the MF of NGC 188. We used artificial-star tests on our data to determine completeness corrections as \citet{wu07} did. The completeness correction for the whole region and for three different regions of NGC 188 are listed in Table 3 for the BATC $e$ band. In the mass functions we included only the values for which the completeness corrections are 0.5 or higher for all regions in NGC 188. Therefore, limiting magnitude $e=20.0$ is adopted. After correcting for completeness, the mass functions for different regions of NGC 188 were obtained, which are shown in Figure 9.

In Figure 9, the MF break is obvious. The break mass is determined to be near $m_{\rm break}=0.885~M_{\odot}$. The MF break of clusters essentially reflects the internal dynamics of clusters on the mass function and/or some fundamental properties of the initial mass function associated to different conditions in star formation \citep{bonatto05a}. The power-law function $\phi(m) \propto m^{\alpha}$ is used to fit the MF. The fit has been performed using a non-linear least-square fit routine which uses the 1 $\sigma$ Poisson errors as weights. The best fitting power-law functions are shown in Figure 9 with solid lines, and the slopes are listed in Table 4. In the fitting of the core region, we did not use the point which only includes one member star. The slope of MF is flat in the core region and become steeper as the distance increases, which reflects the mass segregation of NGC 188. The overall MF slope $\alpha=-0.76$ in the mass range $0.885~M_{\odot}\leq M \leq 1.125~M_{\odot}$ is much flatter than the universal IMF $\alpha=-2.2\pm0.3$ in the mass range $0.5~M_{\odot}\leq M \leq 1.0~M_{\odot}$ \citep{kroupa01b,kroupa02}. This flat slope reflects the advanced dynamical state of NGC 188. The number of stars for NGC 188 can be estimated by the power-law function. We derived the number of stars in NGC 188 based on the power-law function obtained here to be 5776. Based on the number of stars obtained here, we can estimate the relaxation time of NGC 188. The relaxation time is calculated according to this formula $t_{\rm relax}=Nt_{\rm cross}/(8{\rm ln}N)$, where$~t_{\rm cross}=R/\sigma_{v}$ is the cross time, $N$ is the number of stars, and $\sigma_{v}$ is the velocity dispersion \citep{binney08}. We adopted a typical $\sigma_{v} \approx 3~\rm kms^{-1}$ from \citet{binney98}, and the relaxation time of NGC 188 is $t_{\rm relax}=595.6~\rm Myr$. The ratio of the cluster age to the relaxation time is $\tau={\rm Age}/t_{\rm relax}=12.6$. Therefore, NGC 188 is in the advanced dynamical state. Many number of low mass stars have been evaporated and stripped from this open cluster.

\begin{figure*}[htb]
\center
\centerline{\includegraphics[width=0.5\textwidth]{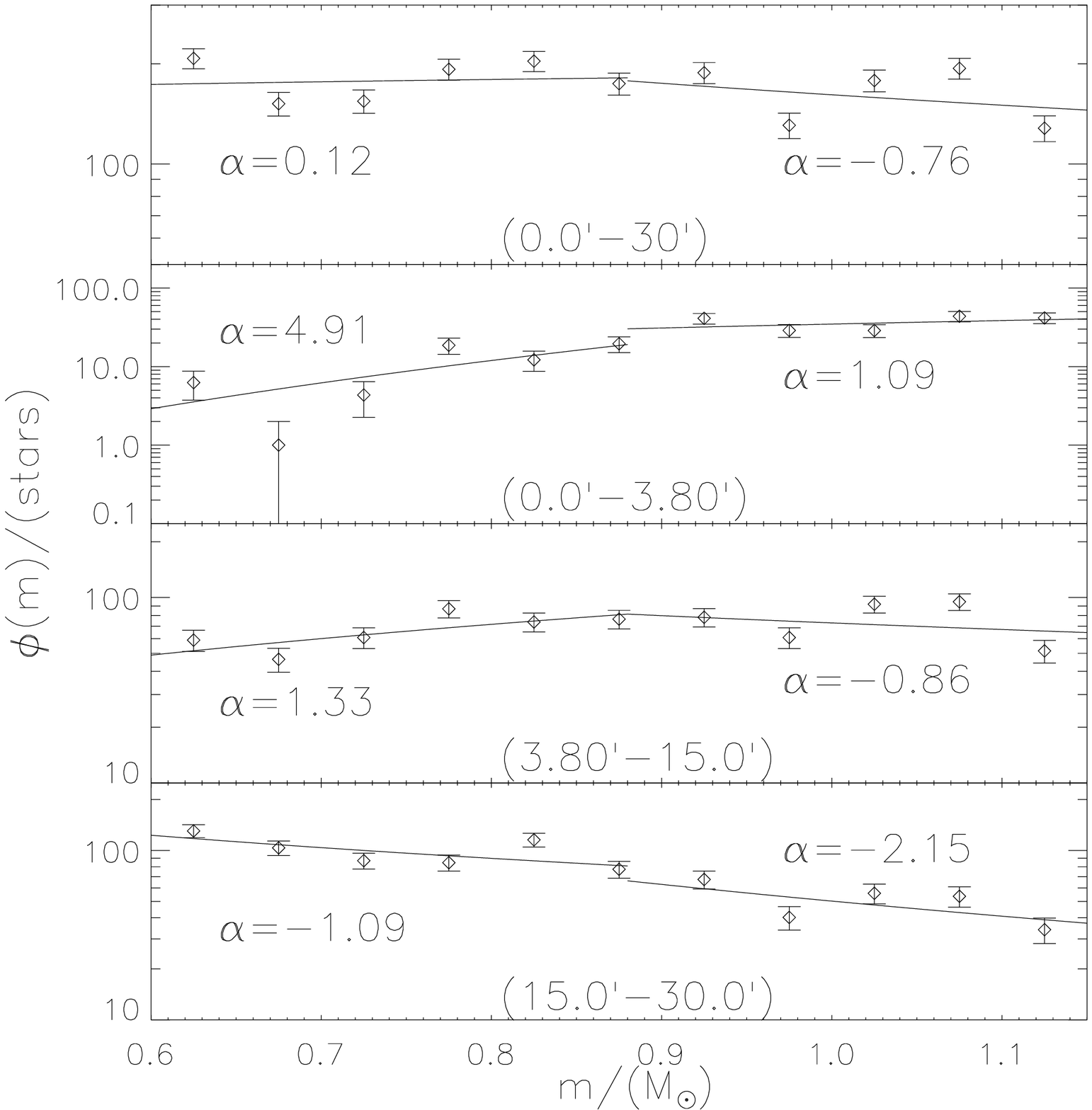}}
\caption{Mass functions in different spatial regions of NGC 188. Error bars represent the 1 $\sigma$ Poisson errors. Solid lines are the best fitting power-law functions. The spatial range and the slope of power-law function are given in each panel. We used stars with $P_{phot}$ greater than $60\%$ to avoid the field-contamination.}\label{fig9}
\end{figure*}

\section{Summary}

In this paper, we present photometry of NGC 188 in 15 intermediate-band filters of the BATC system. By fitting the theoretical isochrones to the observational data points, we determined three fundamental parameters for NGC 188: a reddening of $E(B-V)=0.036\pm0.010$, an age of $t=7.5\pm0.5$ Gyr, and a distance modulus of $(m-M)_{0}=11.17\pm0.08$.

Based on the SED fitting method developed by \citet{wu06}, we obtained a sample of NGC 188 member stars consisting of 2131 stars with $P_{\rm phot}$ greater than $50\%$. We determined the radial surface density profile of NGC 188 using this sample. By fitting the empirical density law of \citet{king62} to our radial surface density profile, the structural parameters of NGC 188 are derived: a core radius of $R_{c}=3.80'$, a tidal radius of $R_{t}=44.78'$, and a concentration parameter of $C_{0}=1.07$.

We studied the details of mass function of NGC 188 based on the photometry which are obtained in a large field of view and deep observation. The MF is fitted with a power-law function, and the slopes of mass functions in different regions are derived. The MF slopes flatten from the outskirts to the inner regions which indicates mass segregation in NGC 188. The MF break in NGC 188 is observed, and the break mass is $m_{\rm break}=0.885~M_{\odot}$. We also derived the number of stars and the relaxation time. The mass segregation and the large ratio of the cluster age to the relaxation time indicate that NGC 188 has gone strong dynamical evolution.

\acknowledgments
We would like to thank the anonymous referee for providing rapid and thoughtful report that helped improve the original manuscript greatly. This study has been supported by the National Basic Research Program of China (973 Program), No. 2014CB845702 and 2014CB845704, and the Chinese National Natural Science Foundation through grants 11373035, 11433005, 11373033, 11203034, 11203031, 11303038, 11303043, 11073032, 11003021, and 11173016.

\clearpage
\setcounter{table}{0}
\begin{sidewaystable}
\begin{center}
\small
\caption{Literature estimates of fundamental parameters for NGC 188} \vspace{0.0mm} \label{table1}
\tabcolsep=2.5pt
\begin{tabular}{ccccccc}
\hline\hline
Reference & Distance Modulus & Distance & Age & $E(B-V)$ & $\rm [Fe/H]$ & Technique \\
 & & (pc) & (Gyr) & (mag) & &\\
\hline
Fornal et al.(2007) & $11.23 \pm 0.14$ & $1700 \pm 100$ & $7.5 \pm 0.7$ & $0.025 \pm 0.005$ & $0.00$(adopted) & $u^{'}g^{'}r^{'}i^{'}z{'}$ photometry,Girardi isochrone fits\\
Meibom et al.(2009) & $11.24 \pm 0.09$ & $1770 \pm 75$ & $6.2 \pm 0.2$ & $0.087$ (adopted) & $-0.1-0.0$(adopted)& Analysis of eclipsing binaries\\
Friel et al.(2010)  & ... & ... & ... & ... &  $0.12 \pm 0.02$ & High-resolution spectral analysis of red giants\\
Jacobson et al.(2011) & ... & ... & ... & ... & $-0.03 \pm 0.04$ & High-resolution spectral analysis\\
This work & $11.21 \pm 0.08$ & $1714 \pm 64$ & $7.5\pm 0.5$ & $0.036\pm0.010$ & $0.12 $(adopted) & BATC multi-color photometry, isochrone fits\\
\hline
\end{tabular}
\end{center}
\end{sidewaystable}
\begin{table}
\centering
\caption{Parameters of 15 BATC filters and statistics of our observations} \vspace{0.0mm} \label{table2}
\begin{tabular}{cccccc}
\hline\hline
\tabletypesize{\scriptsize}
Number & Filter Name& $\lambda _{\rm eff}$ & FWHM & Exposure & Number of Images\\
 & & (\AA) & (\AA) & (s) & \\
\hline
1  & $a$ & $3360$ & $360$  & $4800$   & $7$ \\
2  & $b$ & $3890$ & $340$  & $11010$  & $18$ \\
3  & $c$ & $4194$ & $309$  & $7020$  & $8$ \\
4  & $d$ & $4540$ & $332$  & $6840$  & $7$ \\
5  & $e$ & $4925$ & $374$  & $23940$  & $24$ \\
6  & $f$ & $5267$ & $344$  & $11709$  & $28$ \\
7  & $g$ & $5790$ & $289$  & $7620$  & $14$ \\
8  & $h$ & $6074$ & $308$  & $6780$  & $9$ \\
9  & $i$ & $6656$ & $491$  & $5800$  & $16$ \\
10 & $j$ & $7057$ & $238$  & $7080$  & $10$ \\
11 & $k$ & $7546$ & $192$  & $9300$  & $11$ \\
12 & $m$ & $8023$ & $255$  & $12780$  & $17$ \\
13 & $n$ & $8484$ & $167$  & $12780$  & $10$ \\
14 & $o$ & $9182$ & $247$  & $17985$  & $15$ \\
15 & $p$ & $9739$ & $275$  & $19800$  & $19$ \\
\hline
\end{tabular}
\end{table}

\begin{table}
\centering
\caption{Completeness analysis results for NGC 188} \vspace{0.0mm} \label{table3}
\begin{tabular}{ccccc}
\hline\hline
\tabletypesize{\scriptsize}
$\Delta e$ & $0.0' \leq r \leq 30'$ & $0.0' \leq r < 3.80'$ & $3.80' \leq r < 15$ & $15 \leq r \leq 30$ \\
\hline
$13-14$  & 0.91 & 0.37 & 0.92 & 0.91  \\
$14-15$  & 0.92 & 0.56 & 0.93 & 0.92  \\
$15-16$  & 0.96 & 0.96 & 0.97 & 0.97  \\
$16-17$  & 0.91 & 0.90 & 0.92 & 0.92  \\
$17-18$  & 0.83 & 0.82 & 0.84 & 0.84  \\
$18-19$  & 0.68 & 0.69 & 0.69 & 0.69  \\
$19-20$  & 0.53 & 0.64 & 0.56 & 0.56  \\
$20-21$  & 0.13 & 0.47 & 0.14 & 0.14  \\
\hline
\end{tabular}
\end{table}

\begin{table}
\centering
\caption{Fitted parameters for mass functions of NGC 188} \vspace{0.0mm} \label{table4}
\begin{tabular}{ccc}
\hline\hline
\tabletypesize{\scriptsize}
Distance $r$ & $\alpha~(m < m_{\rm break})$ & $\alpha~(m>m_{\rm break})$\\
$(\rm arcmin)$ & $0.6-0.885$ & $0.885-1.125$ \\
\hline
$0.0-30.0$  & $ 0.12$ & $-0.76$ \\
$0.0-3.80$  & $ 4.91$ & $ 1.09$ \\
$3.80-15.0$ & $ 1.33$ & $-0.86$ \\
$15.0-30.0$ & $-1.09$ & $-2.15$ \\
\hline
\end{tabular}
\end{table}

\begin{thebibliography}{}

\bibitem[Binney \& Merrifield(1998)]{binney98}Binney, J., \& Merrifield, M. 1998, in Galactic Astronomy (Princeton, NJ: Princeton University Press), Princeton serise in astrophysis, QB857.B522

\bibitem[Binney \& Tremaine(2008)]{binney08}Binney, J., \& Tremaine, S. 2008, in Galactic Dynamics (Princeton, NJ: Princeton University Press)

\bibitem[Bertelli et al.(1994)]{bertelli94}Bertelli, G., Bressan, A., Chiosi, C., Fagotto, F., \& Nasi, E. 1994, A\&AS, 106, 275

\bibitem[Bonatto \& Bica(2005)]{bonatto05a}Bonatto, C., \&  Bica, E. 2005, A\&A, 437, 483

\bibitem[Bonatto et al.(2005)]{bonatto05b}Bonatto, C., Bica, E., \& Santos. J. F. C. 2005, A\&A, 433, 917

\bibitem[Bonatto et al.(2006)]{bonatto06} Bonatto, C., Kerber, L. O., Bica, E., \& Santiago, B. X. 2006, A\&A, 446, 121

\bibitem[Carraro et al.(1994)]{carraro94}Carraro, G., Chiosi, C., Bressan, A., \& Bertelli, G. 1994, A\&A, 103, 375

\bibitem[Chabrier(2001)]{chabrier01} Chabrier, G. 2001, ApJ, 554, 1274

\bibitem[Chumak et al.(2010)]{chumak10} Chumak, Y. O., Platais, I., McLaughlin, D. E., Rastorguev, A. S., \& Chumak, O. V. 2010, MNRAS, 402, 1841

\bibitem[Chen(2000)]{chen00}Chen, A. 2000, Ph.D. thesis, Natl. Cent. Univ.

\bibitem[Chen et al.(2003)]{chen03} Chen, L., Hou, J.-L., \& Wang, J.-J. 2003, AJ, 125, 1397

\bibitem[Fan et al.(1996)]{fan96}Fan, X., Burstein, D., Chen, J.-S., et al. 1996, AJ, 112, 628

\bibitem[Fan et al.(2009)]{fan09}Fan, Z., Ma, J., \& Zhou, X. 2009, Res. Astron. Astrophys., 9, 993

\bibitem[Fornal et al.(2007)]{fornal07}Fornal, B., Tucker, D. L., Smith, J. A., Allam, S. S., Rider, C. J., \& Sung, H. 2007, AJ, 133, 1409

\bibitem[Friel et al.(2010)]{friel10}Friel, E. D., Jacobson, H. R., \& Pilachowski, C. A. 2010, AJ, 139, 1942

\bibitem[Girardi et al.(2000)]{Girardi00}Girardi, L., Bressan, A., Bertelli, G., \& Chiosi, C. 2000, A\&AS, 141, 371

\bibitem[Girardi et al.(2002)]{Girardi02}Girardi, L., Bertelli, G., Bressan, A., et al. 2002, A\&A, 391, 195

\bibitem[Girardi et al.(2008)]{Girardi08}Girardi, L., Dalcanton, J., Williams, B., et al. 2008, PASP, 120, 583

\bibitem[Girardi et al.(2004)]{Girardi04}Girardi, L., Grebel, E. K., Odenkirchen, M., \& Chiosi, C. 2004, A\&A, 422, 205

\bibitem[Jacobson et al.(2011)]{jacobson11}Jacobson, H. R., Pilachowski, C. A., \& Friel, E. D. 2011, AJ, 142, 59

\bibitem[Keenan et al.(1973)]{keenan73}Keenan, D. W., Innanen, K. A., \& House, F. C. 1973, AJ, 78, 173

\bibitem[King(1962)]{king62}King, I. 1962, AJ, 67, 471

\bibitem[Kroupa(2001a)]{kroupa01a}Kroupa, P. 2001, ASPC, 243, 387

\bibitem[Kroupa(2001b)]{kroupa01b}Kroupa, P. 2001, MNRAS, 322, 231

\bibitem[Kroupa(2002)]{kroupa02}Kroupa, P. 2002, Science, 295, 82

\bibitem[Marigo et al.(2008)]{Marigo08}Marigo, P., Girardi, L., Bressan, A., Groenewegen, M. A. T., Silva, L., \& Granato, G. L. 2008, A\&A, 482, 883


\bibitem[Meibom et al.(2009)]{meibom09}Meibom, S., Grundahl, F., Clausen, J. V., et al. 2009, AJ, 137, 5086

\bibitem[Platais et al.(2003)]{platais03} Platais, I., Kozhurina-Platais, V., Mathieu, R., Girard, T. M., \& van Altena, W. F. 2003, AJ, 126, 2922

\bibitem[Sandage(1962)]{sandage62} Sandage, A. 1962, ApJ, 135, 333

\bibitem[Sarajedini et al.(1999)]{sara99}Sarajedini, A., von Hippel, T., Kozhurina-Platais, V., \& Demarque, P. 1999, AJ, 118, 2894

\bibitem[Schlegel et al.(1998)]{schl98}Schlegel, D. J., Finkbeiner, D. P., \& Davis, M. 1998, ApJ, 500, 525

\bibitem[Vandenberg \& Stetson(2004)]{vandenberg04} Vandenberg, D. A., \& Stetson, P. B. 2004, PASP, 116, 997

\bibitem[Worthey \& Jowett(2003)]{worthey03}Worthey, G., \& Jowett, K. J. 2003, PASP, 115, 96

\bibitem[Wu et al.(2005)]{wu05}Wu, Z.-Y., Zhou, X., Ma, J., Jiang, Z.-J., \& Chen, J.-S. 2005, PASP, 117, 32

\bibitem[Wu et al.(2006)]{wu06}Wu, Z.-Y., Zhou, X., Ma, J., Jiang, Z.-J., \& Chen, J.-S. 2006, PASP, 118, 1104

\bibitem[Wu et al.(2007)]{wu07}Wu, Z.-Y., Zhou, X., Ma, J., Jiang, Z.-J., Chen, J.-S., \& Wu, J.-H. 2007, AJ, 133, 2061

\bibitem[Yan et al.(2000)]{yan00} Yan, H., Burstein, D., Fan, X.-H., et al. 2000, PASP, 112, 691

\bibitem[Zhou et al.(2001)]{zhou01} Zhou, X., Jiang, Z.-J., Xue, S.-J., Wu, H., Ma, J., \& Chen, J.-S. 2001, Chinese Astron. Astrophys., 1, 372

\bibitem[Zhou et al.(2003)]{zhou03} Zhou, X., Jiang, Z.-J., Ma, J., et al. 2003, A\&A, 397, 361
\end{thebibliography}
\end{document}